\documentclass[
 reprint,
 amsmath,amssymb,
 superscriptaddress,
]{revtex4-1}

\usepackage[all,defaultlines=2]{nowidow} 

\usepackage[section]{placeins} 
\usepackage{color} 
\usepackage{cancel} 
\usepackage{lineno} 
\usepackage[utf8]{inputenc} 
\usepackage{natbib} 
\usepackage{csquotes} 
\usepackage[caption = false]{subfig} 
\usepackage{graphicx}
\usepackage{dcolumn}
\usepackage{bm}
\usepackage{textcomp}
\usepackage{upgreek}
\usepackage{hyperref} 
\hypersetup{colorlinks=true, linkcolor=blue, filecolor=magenta, urlcolor=blue}  
\usepackage{float}

\begin{document}

\title{Thermo-mechanical pain:\\the signaling role of heat dissipation in biological tissues}

\author{Tom Vincent-Dospital}
\email{vincentdospitalt@unistra.fr}
\affiliation{Université de Strasbourg, CNRS, ITES UMR 7063, Strasbourg F-67084, France}
\affiliation{SFF Porelab, The Njord Centre, Department of physics, University of Oslo, Norway}

\author{Renaud Toussaint}
\email{renaud.toussaint@unistra.fr\\}
\affiliation{Université de Strasbourg, CNRS, ITES UMR 7063, Strasbourg F-67084, France}
\affiliation{SFF Porelab, The Njord Centre, Department of physics, University of Oslo, Norway}

 
\begin{abstract}
         \textbf{Abstract}: Mechanical algesia is an important process for the preservation of living organisms, allowing potentially life-saving reflexes or decisions when given body parts are stressed. Yet, its various underlying mechanisms remain to be fully unravelled. Here, we quantitatively discuss how the detection of painful mechanical stimuli by the human central nervous system may, partly, rely on thermal measurements. Indeed, most fractures in a body, including microscopic ones, release some heat, which diffuses in the surrounding tissues. Through this physical process, the thermo-sensitive TRP proteins, that translate abnormal temperatures into action potentials, shall be sensitive to damaging mechanical inputs. The implication of these polymodal receptors in mechanical algesia has been regularly reported, and we here provide a physical explanation for the coupling between thermal and mechanical pain. In particular, in the human skin, we show how the neighbouring neurites of a broken collagen fiber can undergo a sudden thermal elevation that ranges from a fraction to tens of degrees. As this theoretical temperature anomaly lies in the sensibility range of the TRPV3 and TRPV1 cation channels, known to trigger action potentials in the neural system, a degree of mechanical pain can hence be generated.
\end{abstract}
 
\maketitle

\section*{Introduction:\\on rupture and energy dissipation}

The growth of mechanical damages through a body is an irreversible thermodynamic process\,\cite{RICE1978}. Indeed, when a fracture progresses by a given surface unit, it dissipates a specific amount of energy, that is referred to, by rupture physicists, as the energy release rate, expressed in J\,m\textsuperscript{-2}. In most engineering materials (e.g., \cite{G_review}), this quantity, denoted $G$, is well studied, since it characterizes the loading necessary for a crack to propagate\,\cite{Griffith1921}. For instance, it is in the order of $10$\,J\,m\textsuperscript{-2} in weak glasses\,\cite{glass_creep} and can reach $100$\,kJ\,m\textsuperscript{-2} in the strongest media, as titanium\,\cite{tita_creep} or steel\,\cite{stainlsteel_creep}.
When it comes to biological tissues, this energy release rate can also be estimated, and was notably measured to be about $G\sim2000$\,J\,m\textsuperscript{-2} in the human hand skin\,\cite{skinG}. An important question, then, is how is this dissipated mechanical energy received and felt by the human body?\\
Most generally, there are many possible ways for it to be transformed, ranging from its storage as surface potential energy on the walls of the new fractures\,\cite{rice_surface} to its emission to the far field as mechanical\,\cite{crackwaves} or electromagnetic\,\cite{Bouchaud2012} waves, that is, sound and luminescence. It was, in particular, shown that a significant part of the mechanical input is converted into heat close to the damage\,\cite{RiceLevy,Fuller1975,ToussaintSoft,TVD2}, as the rupture of stretched atomic and molecular bonds is prone to generate a local and incoherent -thermal- atomic motion.
The related elevations in temperature have been measured in various synthetic solids (e.g., \cite{Fuller1975,ToussaintSoft,thermal_damage}), and are believed to be more than only a side effect of the fracturing process. Indeed, from its positive feedback on the dynamics of rupture, it was pointed out as a likely cause for the brittleness of matter\,\cite{carbonePersson, TVD1} and for the instability of some seismic faults\,\cite{HeatWeak,pressur2005,SulemCarbo}.\\
We here propose that, in the human body, this damage-induced heat is to be sensed by the neuronal network, and may hence explain a degree of coupling between the thermal and mechanical pain, which has been regularly suspected (e.g., \cite{PainTRP,TRVP1,painless_sensation}).

\section{Thermo-mechanical nociception}

\begin{figure}
\centering
  \includegraphics[width=1\linewidth]{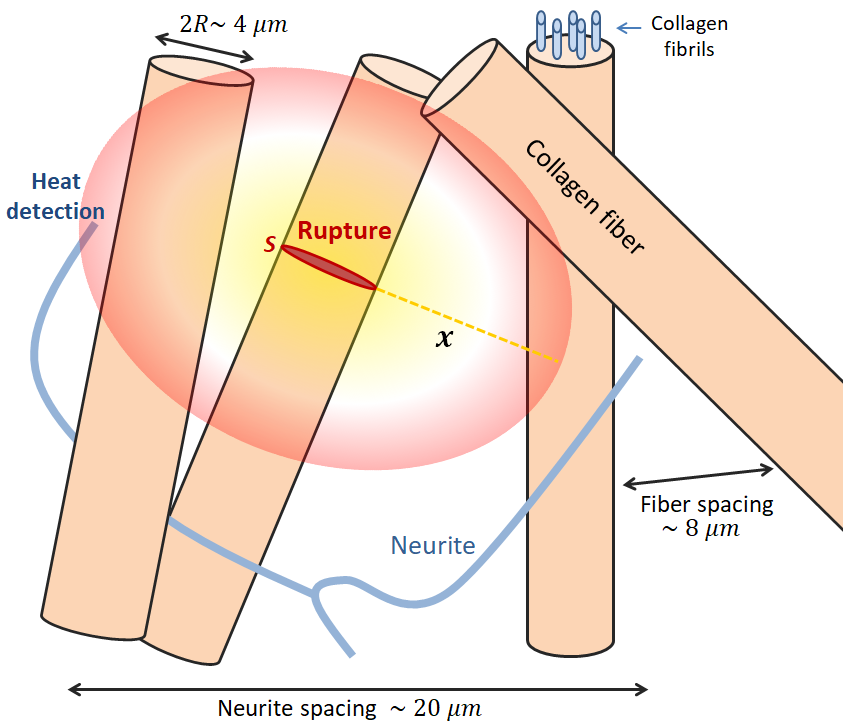}
  \caption{Damage induced heat dissipation in the human skin. A collagen fiber is supposed to brutally snap, after which heat progressively diffuses to the surrounding molecules, as per Eq\,(\ref{green}). If significant enough, this heat migh be detected by the TRPs proteins of the surrounding cells and neurites. In this figure, the schematic collagen and neurites geometries are inspired from \cite{collagen} and \cite{neurite_density_ini}.}
  \label{fig:skin}
\end{figure}
The perception of pain (i.e., nociception or algesia) arises from the bio-electrical signals (referred to as action potentials) that sensory neurons send from the aggressed body part to the nervous system (e.g,\,\cite{innervation}). To initiate such messages, the dolorous inputs, being mechanical, thermal or chemical, need to be converted accordingly, at the surface of sensory neurites (i.e., the extensions of neurons cell bodies).
\\We will here focus, as an example, on the nociception in the human skin. The TRPs proteins (Transient Receptor Potential cation channels) are notably believed to be responsible for the reporting of the temperature of the skin, and that of other body parts, to the nervous system\,\cite{PainTRP,skinTRP}. In particular, TRPV3 send action potentials between $30$ and $40^\circ$C, with an activation intensity that is gradual with temperature\,\cite{TRVP3_grad1,TRVP3_grad2}, leading to a harmless perception of warmth. The feel of a more intense, potentially more noxious, heat occurs when TRPV1 is activated, at higher skin temperatures above $43^\circ$C. Other TRP proteins manage the detection of a cool skin, such as TRPM8 which activates below $30^\circ$C. The physico-chemical mechanism to translate heat into current is, in all cases, believed to be a temperature dependant shifts in the TRPs voltage-dependent activation curves\,\cite{voltage_gating}, that is, in their ability to pass ions through a neuron membrane depending on the balance of charge on each side of this membrane.\\
The role of the TRP proteins is, however, not limited to thermal sensing, and some are known to be sensitive to chemical aggression, responding for instance to abnormal pH, to capsaïcine (i.e., the component of chilli pepper that is felt as hot), to menthol (that is felt as cold), or to arachnid acids\,\cite{PainTRP,TRVP1}. They are hence often referred to as polymodal nociceptors. Similarly, a growing suspicion seems to have risen that these sensors could also be involved in the feeling of mechanical pain\,\cite{PainTRP,TRVP1,painless_sensation}. While the latter is complex, and shall rely on many types of nociceptors other than the TRPs\,\cite{mechano_pain}, the detection of thermal and mechanical inputs has indeed been shown to be somewhat coupled. In particular, the pain threshold of human subjects was reported to be a decreasing function of the ambient temperature\,\cite{coupled_pain}, and, in rats, the drug-induced inhibition of TRPV1 and TRPV3 has proven to reduce mechanical hyperalgesia\,\cite{TRPV1_mechano,TRPV1_mechano2,TRPV3_mechano} (i.e., the increased sensibility to mechanical pain after a first stimulus).
We suggest that this apparent coupling may be explained by the actual (physical) coupling between mechanical damage and heat dissipation. It is for instance well known that burns can be induced by friction on the skin (e.g., \cite{friction_burn}), due to the heat that is there generated. Similarly, a micro-crack of the epidermis or the dermis, that is caused by some mechanical input, is to release some heat in the surrounding tissues, and this heat may well be detected by the skin thermal nociceptors, as illustrated in\,Fig.\,\ref{fig:skin}.

\section{Temperature elevation around a broken collagen fiber}

\begin{figure*}
  \includegraphics[width=1\linewidth]{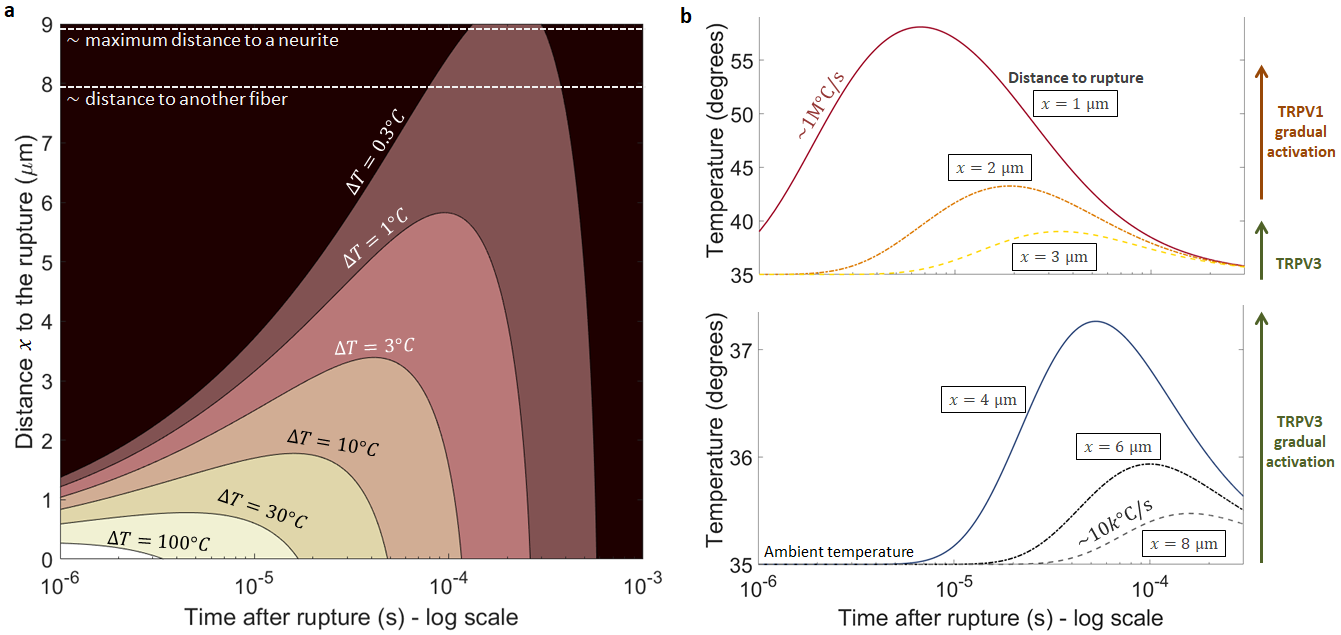}
  \caption{(a): Temperature elevation $\Delta T$ close to a fractured collagen fiber, as a function of the distances in space and time to the rupture event, as predicted by Eq.\,(\ref{green}). (b): Local skin temperature felt at various distances and times from the fractured fiber. Each plots corresponds to a horizontal section of inset (a), to which an ambient skin temperature $T_0=35^\circ$C was added. For readability, two graphs with different temperature scales are shown. (Top): temperature at a distance $x=1$, $2$ and $3\,\upmu$m from the fracture. (Bottom): temperature at $x=4$, $6$ and $8\,\upmu$m. An approximate maximal distance between a neurite and the surface of a collagen fiber is about $9\,\upmu$m, as we developed in the core text. The vertical arrows show the domains of increasing activation of the TRPV1\,\cite{TRVP1} (strong heat) and TRPV3\,\cite{TRVP3_grad1} (warm feeling) protein channels at the surface of neurites.}
  \label{fig:hot_skin_calc}
\end{figure*}
\begin{figure*}
  \includegraphics[width=1\linewidth]{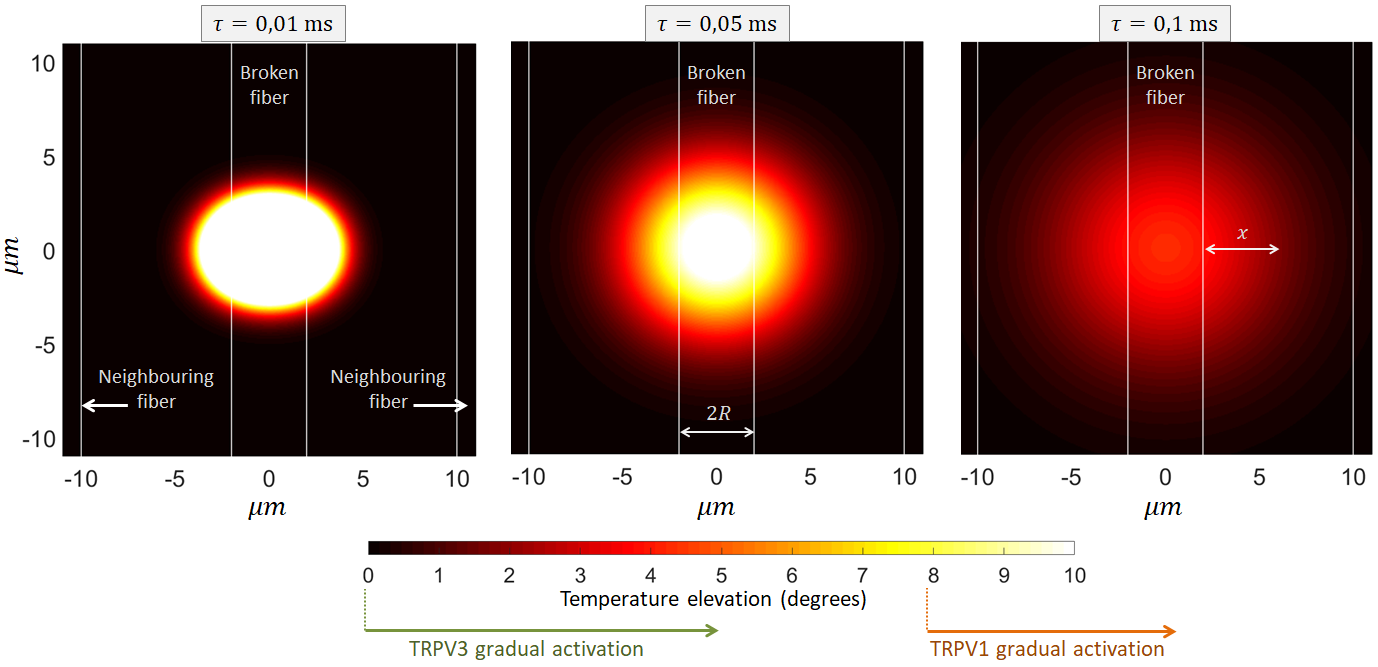}
  \caption{Modelled temperature anomaly ($\Delta T$) around a brutally broken collagen fiber, as per Eq.\,(\ref{green}). Three times $\tau$ after the rupture are shown in chronological order. The color scale is saturated at $10^\circ$C for readability. In each frame, the volume integral $\iiint C\Delta T\mathrm{d}v=\pi R^2G$ is conserved. The white vertical lines mark the border of the broken fiber and of its closest neighbours\,\cite{collagen}. Each map is a cross-section cutting though the central fiber center and has an area of about $1/\rho_n$, where $\rho_n$ is a typical neurite density in the human skin\,\cite{neurite_density_ini}. Thus, at least one neurite is likely to be present on the displayed surface. The two arrows below the color bar represents the domains of activation of the TRPV3 and TRPV1 channels, when assuming a background temperature of $35^\circ$C.}
  \label{fig:skin2}
\end{figure*}

As a major structural cutaneous constituent, let us consider a collagen fiber, which has a typical radius\,\cite{collagen} $R\sim2\,\upmu$m. While it is itself composed of many fibrils and proteins\,\cite{collagen_struc}, the failure of this unit is likely a characteristic step\,\cite{fiber_tissue} in any wider damage of the surrounding matter, as it is the case for engineered fibrous materials (e.g., \cite{ToussaintSoft}). As per the energy release rate $G$ of skin\,\cite{skinG}, the rupture of this fiber shall dissipate an energy $\pi R^2 G\sim25$\,nJ, that is here assumed to be mainly converted into heat on the atomic scale, which then diffuses\,\cite{HeatCond} to the surrounding skin molecules. For simplicity, we suppose that the collagen fiber break is brutal, that is, with a rupture velocity comparable to that of sound in skin\,\cite{Freund1972,skin_ultrason}, $V_0\sim1500\,$m\,s\textsuperscript{-1}. We can then, for various times $\tau$ after the fracture that are superior enough to $2R/V_0=2.5\,$ns, compute the temperature rise $\Delta T$ around the damage, by integrating the Fourier heat diffusion kernel\,\cite{HeatCond} over the broken surface $S$:
\begin{equation}
\Delta T= \iint_{S} \mathrm{d}s\, \frac{G\sqrt{C}}{(4\pi\lambda\tau)^{3/2}}\exp\left(\cfrac{-Cr^2}{4 \lambda \tau}\right).
\label{green}
\end{equation}
In this equation, $r$ is the integration distance between a given point where the temperature is computed, and the various infinitesimal heat sources of $S$, that have an elementary surface $\mathrm{d}s$. In addition, the heat conductivity and volumetric heat capacity of skin are respectively denoted $\lambda$ and $C$, whose values are about $\lambda\sim0.4$\,J\,m\textsuperscript{-1}\,s\textsuperscript{-1} K\textsuperscript{-1} and $C\sim4$\,MJ\,K\textsuperscript{-1}\,m\textsuperscript{-3}\,\cite{skin_thermal}.\\
If the rise in temperature described by Eq.\,(\ref{green}) can be captured by the human neuronal system, it could then be treated as mechanical pain. In a healthy skin, the density of neurites in the torso may be estimated\,\cite{neurite_density_ini} to be about $\rho_n\sim2000$\,mm\textsuperscript{-2}, a quantity from which we derive an order of magnitude for the maximum distance between the surface of a broken collagen bundle and that of a neuronal receptor: $1/(2\sqrt{\rho_n})-R\sim9\,\upmu$m. Interestingly, such an approximate maximum distance is similar to the typical gap between the surfaces of two collagen fibers, which was measured\,\cite{collagen} in average to be about $8\,\upmu$m. Thus, if only two contiguous fibers were to break in response to a mechanical stimuli, one of it would, probably, be rather close (that is, in the micrometer range) to a neurite.\\
Following this simple model, we show, in Fig.\,\ref{fig:hot_skin_calc}, the evolution of the temperature $T_0+\Delta T$ predicted by Eq.\,(\ref{green}), at various distances $x$ up to $9\,\upmu$m perpendicularly to the broken fiber surface, $T_0$ being the normal internal skin temperature.  We use $T_0\sim35^\circ$C. Typical surface skin temperatures can indeed be measured (e.g., \cite{skin_surf_temp}) to lie between $30^\circ$C and $34^\circ$C, with an internal one that should be slightly higher, transiting to about $37^\circ$C (e.g., \cite{skin_temp1}). Of course, a lower skin temperature (for instance at extremities) means that a stronger thermal anomaly would be needed for the TRPs threshold to be reached. Additionally to Fig.\,\ref{fig:hot_skin_calc}, we show, in Fig.\,\ref{fig:skin2}, the related spatial temperature maps at three given times $\tau$ after the fracture.\\
Close to the rupture plane, that is, for $x<2\,\upmu$m, modelled temperatures superior to that of the activation of TRPV1 ($\sim43^\circ$C) are quickly reached, in about $10\,\upmu$s. A painful message can thus be triggered. More conservatively, if the thermal transducers are further away from the rupture point ($x=2$ to $9\,\upmu$m), they undergo a temperature elevation of half a degree to a few degrees, about $0.1\,$ms after the damage. While this quantity is not enough to trigger TRPV1, and not vastly outside the range of the normal temperature oscillations of the human skin\,\cite{skin_temp_variations}, it could still be perceived as some abnormally sudden and localised heat by the brain. TRPV3 was indeed shown to be rather sensitive to small temperature changes around the normal body temperature\,\cite{TRVP3_grad1, TRVP3_grad2}, and with a more intense response the faster these changes are\,\cite{TRVP3_grad2}. Here, as shown in Fig.\,\ref{fig:hot_skin_calc}, we expect very high heating rates ranging from $10$\,k$^\circ$C\,s\textsuperscript{-1} to $1$\,M$^\circ$C\,s\textsuperscript{-1}.

\newpage
\section{Discussion}

\subsection{About the skin model simplicity\label{simplicity}}

\begin{figure}
  \includegraphics[width=1\linewidth]{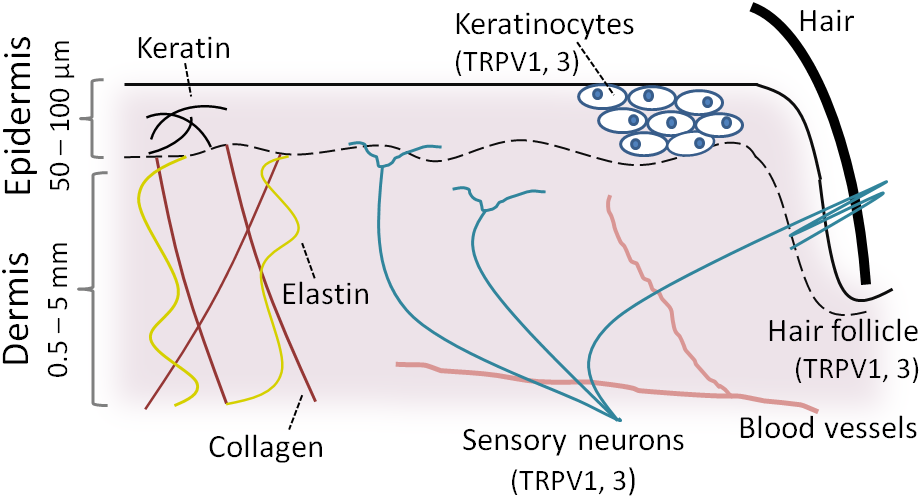}
  \caption{A simple schematic cross-section of the human skin. The main structural proteins of the cutaneous tissues (i.e., collagen, Elastin and Keratin) are notably represented. Without being an exhaustive list, the expression of TRPV1 and TRPV3, reported in various skin parts, is also shown. After\,\cite{skin_anatomy},\,\cite{trp_repartition} and\,\cite{TRVP3_grad2}.}
  \label{fig:schematic_skin}
\end{figure}

Before discussing further the preceding results, and their implication for the feeling of pain, let us acknowledge the simplicity of the model we have considered.
\\In our derivation, we have notably assumed thin, evenly distributed, neurites, that are all thermo-sensitive. This might of course be rather simplified, considering the various types of cutaneous neurons (e.g., \cite{mechano_pain}) and their respective densities in different body parts\,\cite{innervation}. In addition, the expression of the TRPs in the skin (and their role in thermo-sensing) is not limited to its sensory neurons, as they also appear in other cells, notably in the keratinocyte cells\,\cite{PainTRP,extra_TRP1,trp_repartition} of the epidermis. Most of TRPV3 has actually been found in these keratinocyte cells, although it was detected in sensory neurons\,\cite{TRVP3_grad2}.
\\Figure\,\ref{fig:schematic_skin} shows a typical schematics of the human skin, and, by comparison, our conceptual model shown in Fig.\,\ref{fig:skin} is of course only a proxy for the actual cutaneous tissue.
We have notably supposed the thermal properties of skin to be homogeneous. For instance, the TRP proteins reside within the cell membranes, and the heat capacity and conductivity of these lipid bilayers \citep{bilayers_prop} are to slightly differ from the surrounding, aqueous, environment. The extra heat could also be preferably conducted inside the collagen network rather than outside of it. Overall, however, the various components of skin shall have comparable thermal properties. Note also that other conduction laws, more refined than plain Fourier diffusion, may be considered in the modelling of heat transfer in skin (e.g.,\,\cite{skin_nonfourier}).\\
We have, additionally, used a rather homogeneous approach in our mechanical description of skin. In particular, we have looked into the rupture of a single collagen fiber, and it is to be mentioned that the rupture or the frictional contacts of other structural skin proteins, such as keratin and elastin (e.g., see Fig.\,\ref{fig:schematic_skin}), may be also involved in the generation of heat. Yet, collagen accounts for about $80$\% of the dry mass of the fat-free cutaneous tissue\,\cite{skin_anatomy}, and is thus likely involved in most of its significant damages. Most generally, of course, any damage may release some heat burst in the skin, but only that of the toughest (i.e., structural) elements shall generate significant thermal anomalies when rupturing. We have assumed that the energy release rate of collagen was similar to a measured mean skin property\,\cite{skinG} (i.e., $G\sim2000$\,J\,m\textsuperscript{-2}). Such an assumption for $G$ may actually be conservative. Indeed, collagen being a main structural component of skin, it is likely to exhibit a mechanical toughness which is higher than the mean property of the full tissue, which would result in an even larger temperature perturbation than the one estimated here. The considered order of magnitude for $G$ is also in line with the energy release rate of a vast majority of polymeric fibers\,\cite{thin_silk}, and is thus physically coherent. Note however that the internal friction of the collagen fibers (that is, whithout an actual rupture, but which is also prone to release some local heat burst) should play an important role in the cutaneous strength\,\cite{tearskin}. Our focus on the rupture of a simple unit thus remains a simplified view.\\
Overall, the simple approach we followed should lead to reasonable first order estimates of the temperature bursts related to microscopic damages in skin, on which we base the following points of discussion.

\subsection{A thermal anomaly on the edge of detectability}

The modelled damage-induced temperature bursts (i.e.,\,Figs\,\ref{fig:hot_skin_calc} and\,\ref{fig:skin2}) reaching the neurons in less than a millisecond, they could in theory trigger pain reflexes, whose characteristic delays are orders of magnitude bigger, and mainly arise from the two-way travel time of the bio-electrical signals from the neurites to the central nervous system (e.g,\,\cite{innervation}).\\
The time interval during which the temperature elevation holds at a given location is, however, also of importance. In our case, it is of the order of $0.1$\,ms and less (see Fig.\,\ref{fig:hot_skin_calc}), and a question stands on the response time of in situ TRPs proteins. Recent laboratory studies\,\cite{TRP_response2,TRP_response}, applying fast temperature rise to patch clamp samples, indeed suggest that the TRPs reach a steady current emission in times as large as a few milliseconds. It may imply that the temperature signal we have here described should, in practice, be low-pass filtered. While it could stand as a predicament of the theory we propose, a early (transient) response from the proteins channels could still be enough information to be interpreted as pain by the central nervous system.
Furthermore, the response of nociceptors is known to be multifactor, depending for instance in ions concentration and voltage, and the behavior of TRP proteins are known to hold a complex hysteresis. They have notably shown\,\cite{TRP_usedep} an improved response time to temperature jumps after a first excitation. Overall, and while an activation time longer than the millisecond is likely in regard to the current state in the art, the actual responsiveness of in situ channels has not been measured, in particular as a nontrivial amplification of transients can occur in cell signaling (e.g.,\,\cite{noise_ampli}).\\
When it comes to the strength of the temperature anomaly, only part, rather than the whole, of the released energy $G$ could be transformed into heat, leading to an equivalent reduction in our computed temperatures. And, if collagen fibers were to slowly creep rather than brutally snap, more time would be given to the thermal diffusion to evacuate the thus progressively generated heat, so that $\Delta T$ would also be significantly smaller\,\cite{TVD1}.   As an example, if only $50$\% of $G$ was to be dissipated into heat, the local temperature elevation would be twice less than what is shown in Figs.\,\ref{fig:hot_skin_calc} and\,\ref{fig:skin2}, and the threshold for TRPV1 activation would barely be reached, even close to the damage. In practice, this conversion efficiency in organic tissues is not known, and would benefit from some experimental characterisation. However, in a soft polymer (which skin partly is), \citet{TVD2} have shown it to be close to $100$\%, so that this value, although not conservative, is not unsound. \\
The preceding points suggest that the time and amplitude spans of our modelled anomaly are truly at the limit of the acknowledge sensitivity of TRPs. However, we have here only considered a microscopic lesion.   While the rupture of a single fiber is likely representative of the orders of magnitude at stake in this very local phenomenon, larger traumas, in particular if not limited to collagen bundles, could be accompanied with stronger and longer thermal anomalies. There, the spatial extent of the warmed-up neurites may also play some role in the nociception (e.g., \cite{innervation}).

\subsection{Hyperalgesia rather than algesia}

Interestingly, the suspected involvement of TRPs\,\cite{TRPV1_mechano,TRPV1_mechano2,TRPV3_mechano} in mechanical sensibility was reported for hyperalgesia (i.e., the increased sensibility to mechanical pain after a first stimulus) rather than for algesia itself, and we propose that it could be explained by our computed temperatures. Indeed, as we suspect that the energy that is converted into heat lies at the limit of the thermal nociceptors sensibility, the damaged tissues might need to already be inflamed (that is, warmer) for a noxious signal to be generated by a microscopic rupture. Such an effect can, backwardly, be considered as the detection of a sub-threshold background signal (from an inflammation) thanks to a noisy spike (from the rupture of a fiber), and this sort of noise benefice for the nervous system is regularly considered (e.g.,\,\cite{nervous_noise}). Alternatively, but not excludingly, the limitation of the TRPs action to hyperalgesia could arise from the hysteresis in these channels response. It has for instance been shown\,\cite{TRP_response,TRPhysteresis} that TRVP3 needs a first activation at a noxious heat level before responding to temperature changes in a more normal range, which our model covers.

\subsection{On membrane stretching and other mechano-nociception}

We should here restate that the thermo-mechanical pain process that we have described shall certainly not exhaustively account for any sense of mechanical pain. It is rather an explanation to its coupling with the ambiant temperature\,\cite{coupled_pain} and the involvement of TRP proteins. Other mechano-nociceptors are however likely at play (e.g., \cite{mechano_pain}): for instance, the well named Piezo channels, which opening is believed to be directly related to the cell membranes strain, and which could contribute to noxious mechanical sensing\,\cite{piezo2_nox}.\\
Interestingly, as another explanation to the mechanical sensitivity of TRPs, it was proposed\,\cite{forcing_trps} that the stretching of cell membranes could similarly force the opening of their thermal nociceptors. Of course, such a view and that we have developed are not exclusive, as the activation of a channel may, in practice, be polymodal, that is, the thermal responsiveness of a TRP could be enhanced by its abnormal strain.\\
Studies having studied the biothermomechanics of skin also proposed\,\cite{thermal_stress1,thermal_stress2} a pain pathway based on the thermal stress in cutaneous tissues heated with external sources. Indeed, high temperature gradients are prone to cause an inhomogeneous thermal expansion of the skin, and thus, to cause thermal stresses high enough to active its mechano-nociceptors. This view of thermo-mechanical coupling is different than the one we have here discussed. It refers to temperature anomalies leading to stress perturbations, while we suggest that stress-related damages lead to temperature anomalies. It could however stand as a secondary pain mechanism in the framework of our theory, with the rupture-induced thermal bursts causing local stress perturbations. Yet, with a temperature elevation of about $30^\circ$C (see Fig.\,\ref{fig:hot_skin_calc}), a skin thermal expansion coefficient $\epsilon$ of about $10^{-4}$\,K\textsuperscript{-1}\,\cite{thermal_stress2} and a Young modulus $E$ between $10$ and $100$\,MPa\,\cite{thermal_stress2}, such stress perturbation $\epsilon E\Delta T$ should be at least an order of magnitude less than the likely local stress ($\sim 10\,$MPa\,\cite{single_fiber}) having caused a damage in the first place. In our context, it shall thus not be significant.

\subsection{Concluding remarks}

Finally, note that other and similar phenomena than those we have discussed could also be at play in thermo-mechanical sensing. For instance, it was shown (e.g.,\,\cite{tearskin}), that the important toughness of skin is explained by the reorientation and the sliding of collagen fibrils in a stretched skin. The related friction between these collagen units, which occurs before their actual rupture, is also to generate some heat bursts. By contrast, the vasoconstriction in a compressed body would be prone to induce a local cooling\,\cite{blood_temperature}, which could also be sensed.\\
The main concepts we have here discussed, while we have focused on the example of skin, are also general enough to stand for both somatic (that is, related to the skin, tissues and muscles) and visceral (i.e., related to internal organs) pain, in which the TRPs are likely involved, and for which collagen is also a key structural component\,\cite{visceral}.\\
Let us conclude by amusingly pointing out that temperature monitoring has regularly been used by material scientists, including the authors of the present manuscript, to monitor the ongoing damage of engineered solids (e.g., \cite{ToussaintSoft, Bouchaud2012, thermal_damage}), and, in these experiments, we might have unknowingly mimicked our own biology.\\
Such experiments, often focusing on the characterisation of the light emission around propagating fractures, would be a natural next step in the investigation of the hereby presented theory. While mechanical tests on single collagen fibers are possible (e.g.,\,\cite{single_fiber}), it might be challenging to reach, with an infrared camera, the resolution needed to characterise a damage at this fiber level (i.e.,\,a sub-millisecond time and micrometric space resolution). However, studies of larger scale injuries should be more easily accessible, such as the thermo-mechanical studies of the tearing, the cutting, or the puncturing of mammal skin.\\

\subsection*{Acknowledgements and conflicts of interest}

\noindent The authors acknowledge the support of the IRP France-Norway D-FFRACT, and of the Universities of Strasbourg and Oslo. They declare no competing financial interests in the publishing of this work. We also thank Charlotte Kruckow for her insightful suggestions.


\FloatBarrier
\bibliographystyle{unsrtnat}
\bibliography{scr.bib}


\end{document}